\documentclass[aps,prb,superscriptaddress]{revtex4-1}
\pdfoutput=1 
\usepackage[utf8]{inputenc}
\usepackage{natbib}
\usepackage{url}
\usepackage{graphicx}
\usepackage{amsmath}
\usepackage{amsfonts}
\usepackage{caption}
\usepackage{subcaption}
\usepackage{mathtools}
\usepackage[title]{appendix}
\usepackage{bbm}

\begin{document}

\title{Resource allocation for task-level speculative scientific applications: a proof of concept using Parallel Trajectory Splicing}

\author{Andrew Garmon}
\affiliation{Department of Physics and Astronomy, Clemson University, Clemson, SC 29634, USA}
\affiliation{Theoretical Division, Los Alamos National Laboratory, Los Alamos, New Mexico 87545, USA}
\author{Vinay Ramakrishnaiah}
\affiliation{Computer, Computational, and Statistical Sciences Division, Los Alamos National Laboratory, Los Alamos, New Mexico 87545, USA}
\author{Danny Perez}
\affiliation{Theoretical Division, Los Alamos National Laboratory, Los Alamos, New Mexico 87545, USA}
\email{danny_perez@lanl.gov}
\date{\today}

\begin{abstract}
    The constant increase in parallelism available on large-scale distributed computers poses major scalability challenges to many scientific applications. A common strategy to improve scalability is to express the algorithm in terms of independent tasks that can be executed concurrently on a runtime system. In this manuscript, we consider a generalization of this approach where task-level speculation is allowed. In this context, a probability is attached to each task which corresponds to the likelihood that the product of the task will be consumed as part of the calculation. We consider the problem of optimal resource allocation to each of the possible tasks so as too maximize the expected overall computational throughput.  The power of this approach is demonstrated by analyzing its application to Parallel Trajectory Splicing, a massively-parallel long-time-dynamics method for atomistic simulations.
\end{abstract}

\maketitle
\section{Introduction}
As the scale of parallel computers constantly grows, it becomes increasingly difficult for application developers to maintain strong-scalability. For example, on the Summit supercomputer at Oak Ridge National Laboratory, on the order of 100 million operations need to be executed simultaneously in order to fully utilize all processing elements. As the number of processing elements is expected to steeply increase as we approach the exascale era \cite{alexander2020exascale}, it is paramount to develop novel strategies to maximize the amount of parallelism exposed by the applications. A now common programming model in scientific applications is task-based programming, where the execution of the application is factored into tasks of varying granularity that are then scheduled for execution using a runtime system \cite{Robson:2016:RCH:3018814.3018821,bauer2012legion}. This model has proved to be powerful in a range of contexts, and his now deployed in production scientific applications \cite{di2020htr,doi:10.1063/5.0014475,torres2019soleil,jain:isc2016}.
A potentially promising generalization of task-based programming to further performance at very large scales is task-level speculative execution, akin to a distributed-memory version of Thread Level Speculation \cite{10.1145/2400682.2400698,10.1145/2821505}.  In our approach, (coarse) computational tasks are made available for execution before it can be established that they will definitely be used as part of the calculation. If speculations can be made accurately so that the results of most executed tasks are eventually used,  this strategy has the potential to enable higher concurrency, and hence to improve scalability.

This manuscript considers the problem of optimal resource allocation in a speculative task-execution setting where a task usefulness probability, i.e., a probability that the results of a speculatively executed task will be consumed as part of the calculation (hereby referred to as the \textit{task probability}), can be explicitly computed or estimated. This is a rich problem as, in practice,  task probabilities will often be conditioned on the current state of the application, and can therefore dynamically change as execution proceeds. Further, the run time of individual tasks is often much shorter than the run time of the application as a whole. As a result, as tasks complete, freeing-up previously allocated resources, new tasks have to quickly be identified to take their place. Finally, optimal  allocation of resources to tasks (i.e., how much computing resources are dedicated to the execution of each given task) is not only dependent on the individual task probabilities, but is also tied to the distribution of task probabilities of all other tasks that are available for execution.

In the following, we consider such a dynamic setting where tasks are assumed to be preemptable and restartable with a different resource allocation. It then becomes possible to periodically reevaluate the optimal allocation, pausing the execution of all running tasks, and reallocating resources as needed. This can either be done at fixed time intervals, when tasks complete, or when a change of context dictates. With each update, a newly derived optimal allocation can be executed, resuming paused tasks (with a potentially different resource allocation) as well as starting new ones if needed. This pausing and resuming of tasks allows for the optimal allocation to adapt to the dynamic variability of the system, maintaining an optimal (expected) throughput at all times. In the following, we present a generic analysis of this approach as well as a case study to a specific scientific application called Parallel Trajectory Splicing \cite{ParSplice}, which is adapted to a setting where task probabilities can be explicitly estimated.

\section{Previous work}

Modern parallel computing architectures have complex memory hierarchies as well as heterogeneous processors. In order to achieve high performance on such architectures, programming models such as Legion \cite{bauer2012legion} are organized into logical regions that expresses locality and independence of data and tasks. The instances of these logical regions can be assigned to specific memories and processors in the machine during run-time. Similar logical hierarchies are also introduced in OpenMP 5.0 \cite{OpenMP5}, Chapel \cite{chamberlain2007parallel}, Charm++ \cite{kale1993charm++}, etc. for task-based parallelism. These task-based systems are capable of dynamic load balancing for scheduling and mapping tasks for optimal performance on the underlying hardware. Locality-aware parallelism has been well studied in non-speculative systems, and only a select few speculative systems utilizing parallelism via thread-level speculation (TLS) \cite{steffan2000scalable} or hardware transactional memory (HTM) \cite{bobba2007performance} can scale beyond a few nodes. One such system that we came across is described in the work by Jeffrey et al., \cite{jeffrey2016data} where program knowledge is leveraged to provide \textit{spatial hints} to indicate the data that is likely to be accessed by a speculative task. In this work, we adapt and augment this idea and speculatively schedule tasks based on their usefulness in contributing to the overall computations in order to increase throughput.

Many resource allocation strategies have been explored in the context of load balancing to efficiently use the existing hardware. A naïve way to allocate resources is to base it on peak utilization. However, designing a resource allocation strategy based on worst-case needs is not a viable approach as it results in excessive resource estimates. Many   static and dynamic approaches \cite{almeida1995comparative,andonov1993dynamic,morales1995integral,algorithms1988gibbons} have been proposed to distribute the problem pieces optimally over different nodes with an objective of balancing the execution time. However, the issue with most of these optimization problems is the curse of dimensionality as the search space grows exponentially with the size of the problem and the potential impact of emerging hardware, such as smart interconnects \cite{rajamony2011percs,faanes2012cray} with advanced traffic monitoring hardware. Static approaches distributing the load during compile time have limitations as the performance is not only dependent on problem size but also over many dynamic factors. Adaptive resource management techniques \cite{rosu1997adaptive} try to overcome these limitations by dynamically allocating resources to different processes. To provide software support, the MPI-2 standard also introduced dynamic process creation using the MPI\_Comm\_spawn function \cite{balaji2011mpi}. This function enables to create new processes during the program run-time. To mitigate poor resource allocation and load balancing in dynamic MPI spawning, fuzzy scheduling algorithms \cite{moussa2017intelligent} for dynamic processes have been explored.
 
Many control-theory based techniques are also used for adaptive resource allocation that use standard feedback controllers with an auto-regressive prediction model to predict the resource allocation \cite{xu2009grey}. Many resource monitoring, prediction, and allocation strategies have been explored in cloud computing environments \cite{minarolli2014distributed,wei2015towards,ma2016auto}. Solutions including genetic algorithms \cite{tseng2017dynamic}, neural networks, etc. are explored for prediction and allocation of resources in cloud data centers \cite{chen2015resource}. All of these approaches are based on learned behavior from heuristics and do not consider the inherent probabilities of individual tasks at an application level.
 
In any resource allocation problem \cite{morales2000design}, limited resources are to be allocated to a set task to maximize effectiveness. 
Dynamic programming has also been explored in this setting and it can be shown \cite{elmaghraby1993resource,powell2002adaptive,denardo2012dynamic} that the problem can be solved using a simple sequential multi-stage dynamic programming algorithm in $\mathcal{O}(N^2M)$ time. Pipeline based algorithms \cite{gonzalez2003towards,jahn2015runtime} that mimics instruction pipelines within processors have also been attempted, however, most of these approaches have a high communication cost.

\section{Methods}

\subsection{Optimal Throughput}
Consider the problem of allocating resources between a (potentially infinite) number of candidate tasks in a speculative task execution setting on a machine containing $N$ hardware slots on which tasks can be assigned (which can be nodes, cores, GPU, etc).  Tasks can be run in parallel over a certain number of slots $w_i$, in which case, task completion requires an expected time $T(w_i)$. For simplicity, it is assumed that all tasks are computationally equivalent; however, a task-specific $T_\alpha(w_i)$ can be introduced in the derivation below without additional complication.

Each of the candidate tasks are assigned a probability $p_i$ of ultimately being used as part of the overall execution of the calculation, which is abstractly conceived as a workflow that progresses by consuming completed tasks. This probability can either be a rigorously derived value or a heuristic estimate. Note that each task can also be assigned a weight that reflects how much it would contribute to the calculation by scaling the corresponding $p_i$ accordingly to obtain an expected utility. In the following, however the $p$’s will still be referred to as probabilities, for simplicity.

To simplify and accelerate the allocation process (which is important in the context where probabilities are adjusted and resources re-allocated at a high rate), the resource assignment problem is solved in a continuous setting where the $w_i$ are real numbers instead of integers. This enables an extremely efficient solution scheme. These values can then be discretized after optimization is complete, yielding an approximate solution but at a greatly reduced computational cost.  In what follows, it is assumed that the tasks are ordered by decreasing probability. The optimal allocation of resources consists of determining the number of tasks that should be executed, $M$ (i.e., the $M$ tasks with the highest probabilities are selected for execution) as well as the resources assigned to each task, $w_i$.

The objective function to be optimized is the instantaneous expected throughput from the $M$ tasks that are selected for execution:
\begin{center}
    $R(M,\{w_i\})=\sum_i^M{p_i/T(w_i)}$
\end{center}
where $T(w_i)$ is the expected time to complete a task when provided $w_i$ resources. $R(M,\{w_i\})$ measures the expected rate at which useful results are generated for a given allocation $M,\{w_i\}$.  The problem is constrained by requiring that the allocation fully utilizes available resources, so that $\sum_i^M w_i = N$. In pathological cases where there are more resources available than could possibly be used ($N>M*w_\mathrm{max}$, where $w_\mathrm{max}$ is the maximum allocation which a single task can fully utilize, as will be defined below) $N$ is replaced with $M*w_\mathrm{max}$. This constraint can be enforced by introducing a Lagrange multiplier $\lambda$ to the objective function.

\begin{center}
    $R(M,\{w_i\})=\sum_i^M{p_i/T(w_i)}+\lambda(\sum_i w_i-N)$
\end{center}

Extremizing the Lagrangian with respect to $w_i$ and $\lambda$ yields

\begin{center}
    $p_i F(w_i)+\lambda=0$ \\
    $\sum_i w_i=N$ \\
\end{center}
respectively, where the function $F$ is defined $F(w_i) \coloneqq-T^{'}(w_i)/T^2(w_i)$.

Therefore, given an explicit expression for $T(w_i)$, one can invert the function $F$ and obtain an explicit expression for $w_i$:
$w_i=F^{-1}(-\lambda/p_i),$
which depends only on the Lagrange multiplier $\lambda$ and on the task probabilities $p_i$.
At a given value of $M$, the allocation problem is reduced to solving a 1D root-finding problem in $\lambda$, $\sum_i w_i=N$, which yields the values of $\{w_i\}$ that maximizes the throughput for this value of $M$. Note that this formulation can yield $w_i=0$, so that considering the first $M$ tasks in the optimization problem is not guaranteed to allocate resources to all of them. Finally, the optimal number of tasks to consider, $M^*$, is taken to be the value which maximizes the expected throughput over all values of $M$. The allocation problem is therefore reduced to solving two embedded 1D problems, which can be done very efficiently.

\begin{figure}
    \centering
    \includegraphics[width=\linewidth]{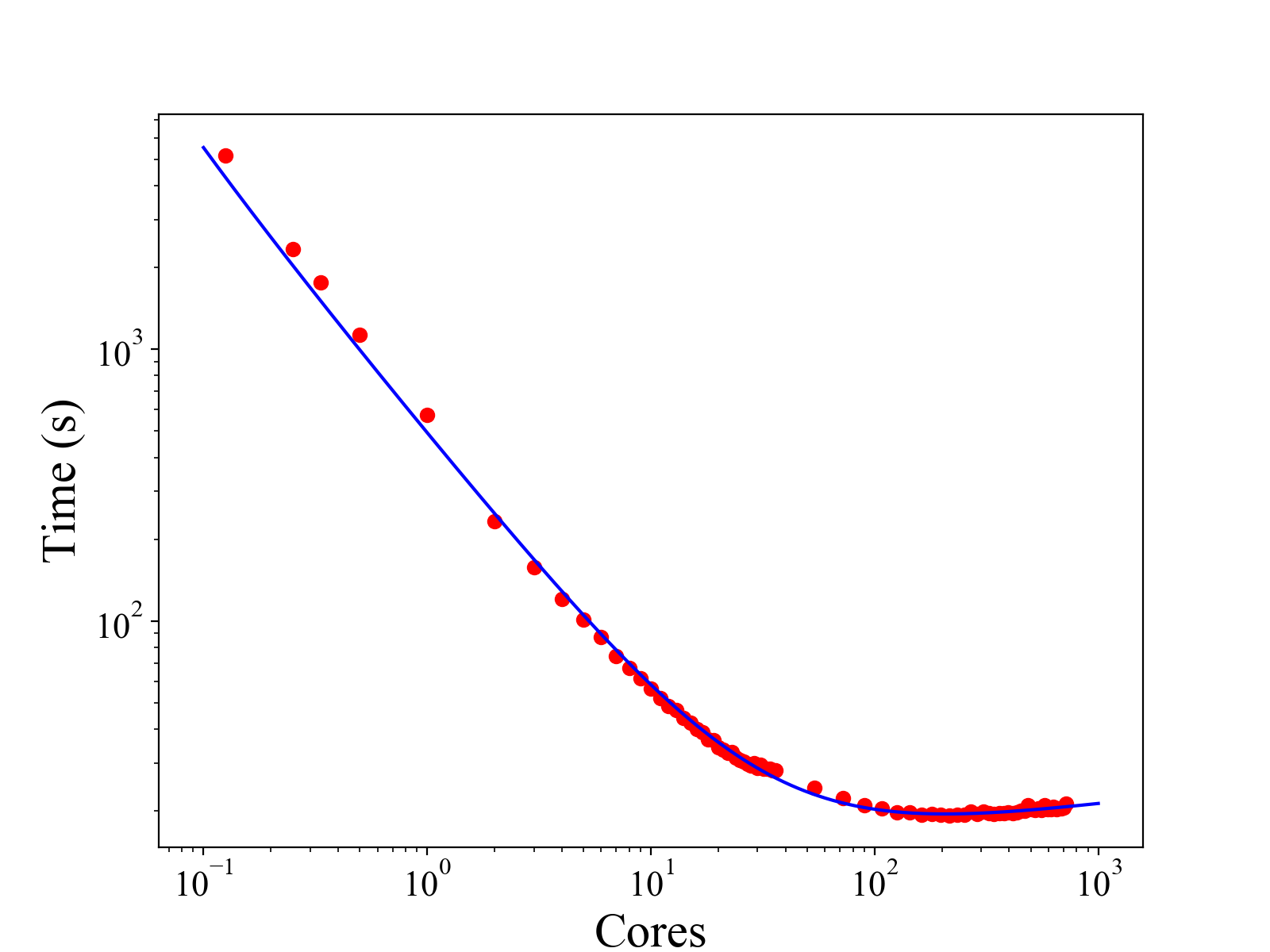}
    \caption{Benchmark analysis of the molecular dynamics code LAMMPS \cite{LAMMPSCODE}. Run times in red were measured for an identical task executed over a varying number of cores. Fractional core counts were obtained via oversubscribing the hardware slots. In blue, a functional form $a+b/x+d\log(gx)+h/x^2$ was fit to the data to produce the invertible function $T(w)$ with coefficients $a=-2.38, b=481.42, d=2.32, g=21.76,$ and $h=7.10$ }
    \label{fig:benchmark}
\end{figure}

In practice, an explicit expression for $T(w_i)$ is obtained by fitting to benchmark results. Benchmarks were carried out on dual sockets Intel Broadwell E5-2695V4 nodes.  In section \ref{ApplicationSection}, an application to parallelized materials simulation is considered. For this work, a benchmark analysis of the molecular dynamics code LAMMPS \cite{LAMMPS,LAMMPSCODE} was conducted as shown in Figure \ref{fig:benchmark}. The function $T(w_i)$ was obtained by running an identical task in parallel over a varying number of cores and recording the time to complete said task. Fractional core counts are obtained when oversubscribing the hardware slots. The recorded times were then fit to the functional form $a+b/x+d\log(gx)+h/x^2$ which was loosely based on Amdahl's law \cite{amdahl1967validity}, adding a heuristic $\log$ term to account for the cost of synchronization and a $1/x^2$ term to provide an oversubscription penalty. The specific functional form is not crucial; other smooth approximations could have been used instead. 

As shown in Figure \ref{fig:benchmark}, $T(w)$ possesses a minimum, after which the time to execute a task begins to increase with increasing resources due to communication and synchronization overheads.  As no optimal resource allocation can include $w$’s in this regime (because a higher throughput could always be obtained with even fewer resources) this branch of the function $T(w)$ is ignored when obtaining $F^{-1}$. The minimum of $T(w)$ therefore defines the maximum allocation ($w_\mathrm{max}$, roughly $200$ cores in this case) which can be fully utilized by a task, and hence a corresponding minimum time in which a task can be completed; here $T(w_\mathrm{max})$ is roughly $19.5$ seconds. This quantity becomes important in conjunction with $T(1)$, the time to complete a task at the maximum parallel efficiency, as their ratio will be shown to correspond to an upper bound of achievable performance gains. In addition to ignoring the $w>w_\mathrm{max}$ branch, the domain of $F$ is restricted to those values of $w$ where $F$ is monotonically decreasing, which is required for the solution to be a maximum of the throughput, in contrast to a minimum. $F$ is therefore invertible so that $F^{-1}$ is well defined.

\subsection{ParSplice}
In the following, the potential benefits of optimal resource allocation in a speculative task execution setting are demonstrated by studying an existing scientific application called Parallel Trajectory Splicing, or ParSplice \cite{ParSplice,perez2017long}. ParSplice is a method in the family of Molecular Dynamics (MD) simulations.
MD numerically integrates the classical equations of motion of atoms using interatomic forces derived from the gradient a user-provided potential that describes the interactions between atoms. MD is broadly used in the computational sciences, with applications to materials science, biology, chemistry, etc. MD is extremely powerful, but also computationally intensive. While domain-decomposition approaches enable the use of massively-parallel computers to extend the simulation length-scales \cite{LAMMPS}, similar approaches do not allow for significant extension in timescales except for very large systems, due to communication and synchronization overhead. Extending timescales instead requires specialized techniques \cite{AKMC,hyperdynamics,TAD,Zamora2020,perez2015parallel}. ParSplice is one such technique where parallelization is carried out in the time domain, thereby avoiding synchronization costs inherent to domain decomposition. It however comes with a tradeoff: instead of generating a trajectory that is continuous in phase space, it produces a discrete state-to-state trajectory, where a state corresponds to a finite volume in the phase space of the problem. States are usually defined to correspond to long-lived metastable topologies of the system (such as the attraction region of deep local energy minima), and so state-to-state trajectories are sequences of transitions between such long-lived states.

\begin{figure}
    \centering
    \includegraphics[width=\linewidth]{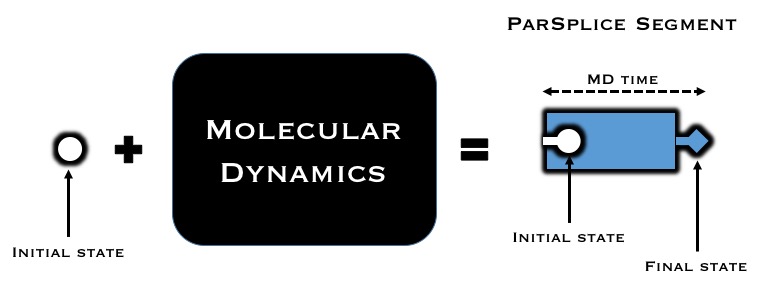}
    \caption{Conceptual illustration of segment generation: An MD trajectory is initialized in some assigned ``circle" state and then dynamically evolved forward in time through MD. After some stopping criteria is met, the final state of the MD trajectory is noted and used to produce a ParSplice ``segment".}
    \label{fig:genSeg}
\end{figure}

\begin{figure}
    \centering
    \includegraphics[width=\linewidth]{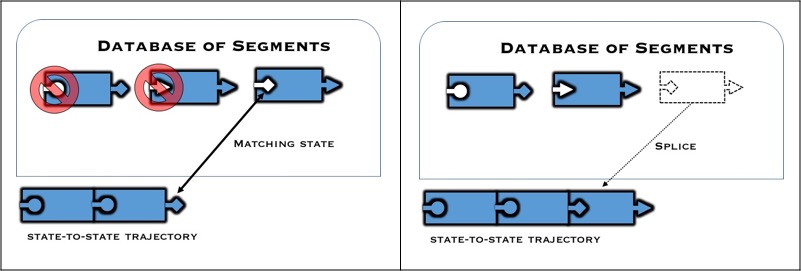}
    \caption{Conceptual illustration of segment splicing: Left panel, only segments which start in the same state that the previous spliced segment ended (here the ``diamond" state) can be spliced. Right panel, splicing a segment involves extracting it from the database and appending it to the state-to-state trajectory.}
    \label{fig:spliceSeg}
\end{figure}

ParSplice works by concurrently and asynchronously generating many short “segments” of MD trajectory in such a way that they can later be spliced together to create a single state-to-state trajectory. Generating a “segment” involves creating an independent realization of the system’s trajectory (by solving a stochastic differential equation) that is initialized in some assigned starting state and evolved through MD until a physics-motivated stopping condition is achieved, after which the final state visited by the trajectory (which may or may not be the same as the starting state) is noted. So, in short, a segment is composed of an initial and a final state, separated by some MD time (see Figure \ref{fig:genSeg}). These segments are then returned to a database where they are stored until they can be spliced. Due to the specially-designed protocol by which segments are produced and stored \cite{aristoff2019generalizing}, any segment in the database can be spliced onto any other so long as it began in the same state that the other finished (see Figure \ref{fig:spliceSeg}). This allows for a single state-to-state trajectory to be formed by extracting individual segments from the database and splicing them onto the end of the trajectory. For further details on how the independent generation and splicing of segments is guaranteed to produce statistically correct state-to-state trajectories, the reader is referred to the original manuscript \cite{ParSplice}.

Because the individual segments are independently produced in parallel, ParSplice can offer a potential wall-clock speedup proportional to the number of MD instances. Achieving this ideal level of parallel efficiency however requires that every segment generated is eventually spliced into the state-to-state trajectory. Therefore, while the accuracy of a trajectory is ensured solely by the independent generation and splicing of segments according to the ParSplice prescriptions, the efficiency of ParSplice is a function of its ability to forecast ahead of the trajectory and assign segments to be generated in those states where they are most likely to be needed. As such, ParSplice follows the speculative execution paradigm discussed above: at any point in time, only one segment is strictly guaranteed to be spliced into the trajectory (a segment that begins in the state where the trajectory currently ends), but one can identify a much larger number of segments that could potentially be spliced in the future.  Towards this goal, ParSplice develops a discrete time Markov Model (MM) on-the-fly from the previously generated segments and uses this model to assign starting states for new segments to be generated. The MM encodes the estimated probability that a segment generated from state $i$ will end in state $j$. In actual simulations, the MM is usually empty at the beginning of the simulation and it is dynamically updated as more segments are generated.

\begin{figure}
    \centering
    \includegraphics[width=\linewidth]{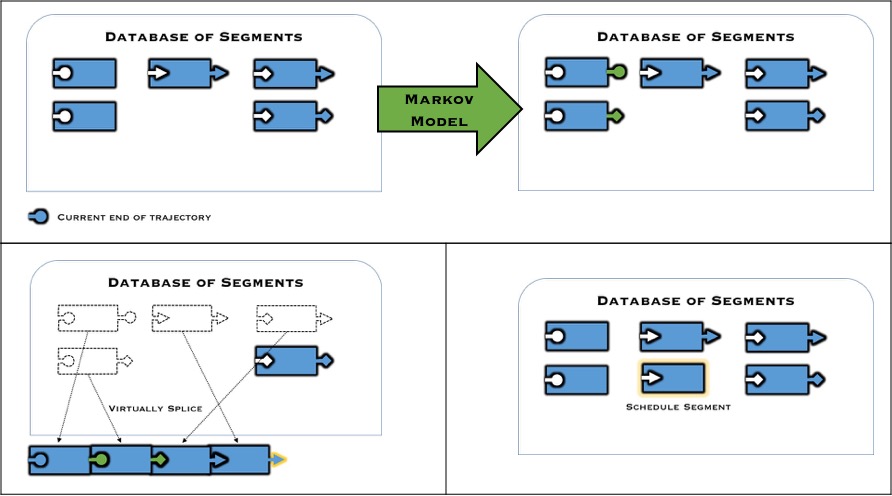}
    \caption{Virtual End (VE) scheduling of segments:
    Top panel, the statistical Markov Model (in green) is used to sample ``virtual" end states (also green) for all pending-segments, speculating on what the database might look like once these pending-segments are completed. Bottom-left panel, segments are then ``virtually" spliced from the speculative database, extending the ``virtual" trajectory as far as possible. Bottom-right panel, a new segment (outlined in yellow) is scheduled  to begin in the state where the ``virtual" trajectory ended. We stress the word ``virtual" here to differentiate from anything actual. All segment manipulation is only carried out as a thought experiment for determining where to generate the next segment.}
    \label{fig:VE}
\end{figure}

The original ParSplice method selects segments for execution through a procedure referred to as virtual end (VE) scheduling. VE  accounts for completed but unspliced segments which are stored in the database as well as those “pending” segments which have been assigned to some computing resources, but have not yet been completed. The process of VE assigning the state in which the next segment should be generated is outlined in Figure \ref{fig:VE}. 1) The MM is used to sample “virtual” endpoints for all of the pending segments, creating a prediction of what the database might look like once all of the pending tasks are completed. 2) It then “virtually” splices from this database-prediction onto the end of the state-to-state trajectory until it runs out of segments to splice. 3) It assigns the next segment to be generated starting in the state where the state-to-state trajectory “virtually” ended. This process is then repeated for next segment state-assignment, and so on until a segment has been assigned to any idle MD instance.
The word ``virtual" is used to denote that this process is not actually manipulating segment endpoints or splicing onto the actual physical trajectory. This process is simply used as a means of forecasting where to assign the next segment and only segments that were actually completed can be spliced into the physical trajectory.

In the present context, an important limitation of the VE procedure is that it samples from the ensemble of possible tasks according to their probabilities, but does not directly give access to the individual task probabilities themselves. 
In order to address this limitation, a new variant of ParSplice is proposed where instead the task probabilities are first explicitly estimated, and then tasks with the largest probabilities are selected for execution. This new variant is referred to as MaxP (maximum probability) scheduling. The derivation of MaxP is based on the formalism of discrete time Markov Chains, and is detailed explicitly in Appendix A. The general concept involves calculating the probability that particular segments will be spliced into the state-to-state trajectory over some finite time horizon, as an average over all paths that the spliced trajectory could take. These probabilities can be computed analytically or approximated via a computationally cheaper Monte Carlo approach. See Appendix A for details. 

The MaxP formulation provides a natural estimate of the task probabilities for each segment that could be generated, i.e, each potential task. It is important to note that the probabilities derived from the MaxP formalism are dependent both on the instantaneous content of the database and on the current end point of the trajectory, as it was the case for the VE variant. The probabilities therefore continually change as the simulation proceeds, which suggests that it might be advantageous to periodically re-adjust/recalculate the probabilities and re-assign resources to tasks so as to maintain an optimal expected throughput. Further, MD is inherently preemtable and restartable: the only information needed to checkpoint and restart a simulation is the list of the current positions and velocities of the atoms. Using this checkpoint, the simulation can be restarted with a different domain decomposition, and hence with a different $w$. The resource allocation approach discussed above is therefore directly applicable to ParSplice-MaxP.

\section{Application}
\label{ApplicationSection}

To gain a better intuition of the solutions resulting from different task probability distributions, and of the potential performance improvements that can be expected by allocating resources based on task probabilities, we first discuss results on various synthetic distributions. More specifically, we focus on the characterization of the instantaneous throughput obtained by optimizing the resource allocation as a function of the characteristics of the task probability distribution. Each of the following distributions were created by drawing random $p_i$ samples from a probability density until a given total $\sum p_i=1000$ was reached. While this process resulted in each synthetic distribution containing a different number of potential tasks, the constrained value of $1000$ ensures that the maximum expected throughput given infinite resources is identical for each distribution, and hence comparisons can be made easily. 

The probability densities from which the synthetic distributions were sampled belonged to one of two generic classes. The first was a delta distribution or composition of two delta distributions from which only particular values of $p_i$ could be sampled. Each composition of delta distributions contained a non-zero peak at $p=1$, corresponding to having a certain number of tasks which are known to be essential (i.e $p=1$), and another non-zero peak at lower $p$, corresponding to a certain number of speculative tasks which are assigned a generic probability. As one would expect to generally have a large number of speculative opportunities, and thus of speculative tasks, the magnitude of the peaks were weighted in favor of the lower probability by a 9:1 ratio. Sampling from these distributions yields a task probability distribution which exhibits a ``step” from the $p=1$ tasks to the speculative probability. In addition to a single delta distribution at $p=1$, which generated a trivial distribution containing only $p=1$ tasks, several composite distributions are analyzed with varying values for the lower-probability speculative tasks.

\begin{figure}
     \centering
     \includegraphics[width=\linewidth]{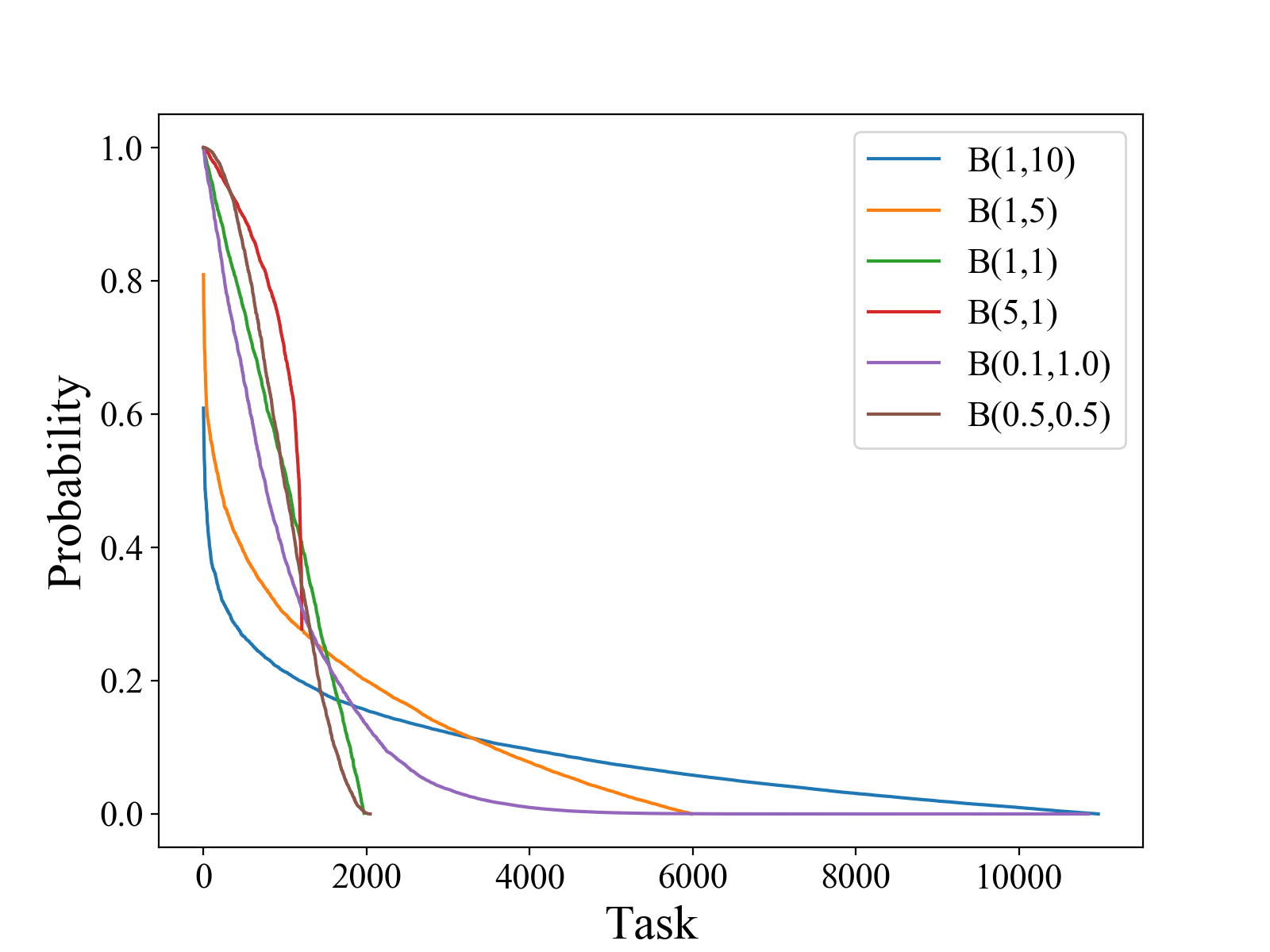}
     \caption{Synthetic task probability distributions sampled from different B$(\alpha,\beta)$ distributions as depicted in the legend.}
     \label{fig:BetaP}
 \end{figure}

The second class of probability densities were beta distributions, B$(\alpha,\beta)$. Adjusting the shape parameters $\alpha$ and $\beta$ allows for the creation of a wide range of different distributions, as illustrated in Figure \ref{fig:BetaP}. Sampling from the continuous probability densities produced nearly continuous task probability distributions capable of spanning the entire $[0,1]$ probability domain. 
This assortment of synthetic task probability distributions provide a reasonable collection for surveying the performance landscape of the proposed optimal resource allocation method.
 
The most important question in practice is whether the effort of deriving and implementing a probability-aware optimal allocation scheme is worthwhile as compared to a naive approach which does not consider the probability of tasks. Such a naive scheme would assign resources in equal sized chunks corresponding to the maximal parallel efficiency, so as to maximize throughput in the non-speculative setting. It would only deviate from this chunk size if the resources available enabled all tasks to be run at maximum parallel efficiency and excess resources remained. In such a case, the constant chunk size allocated to each task would uniformly increase to fully utilize all available resources. Therefore, the naive allocation is to assign each task with $w_\mathrm{const}=\max(1,N/M)$ resources to each task. In the following, it is shown that the increase in throughput due to optimal allocation can in fact be quite substantial. 


\begin{figure}
    \centering
    \begin{subfigure}[b]{.49\textwidth}
         \centering
         \includegraphics[width=\linewidth]{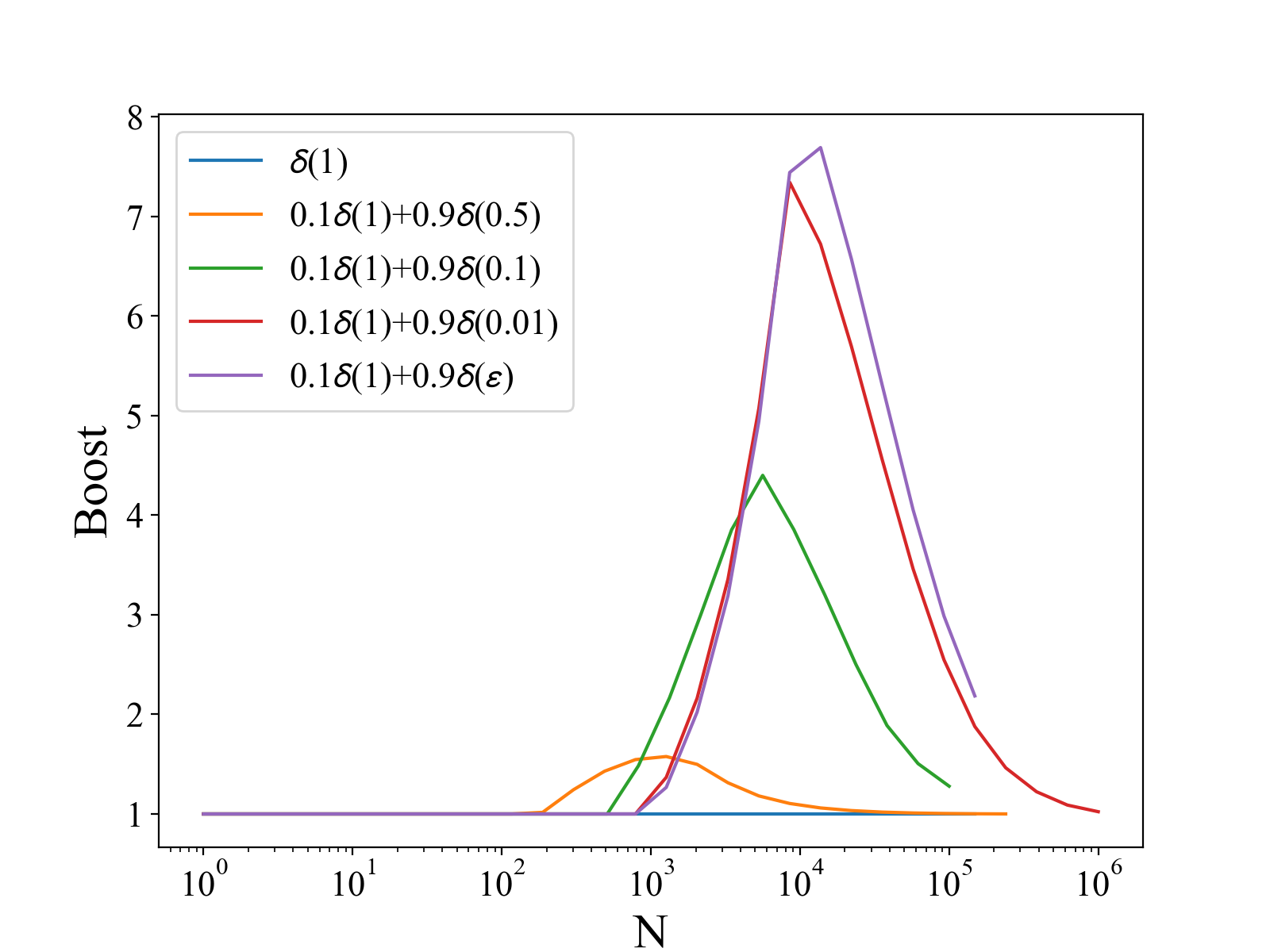}
         \caption{}
         \label{fig:Deltaboost}
     \end{subfigure}
     \hfill
     \begin{subfigure}[b]{.49\textwidth}
         \centering
         \includegraphics[width=\linewidth]{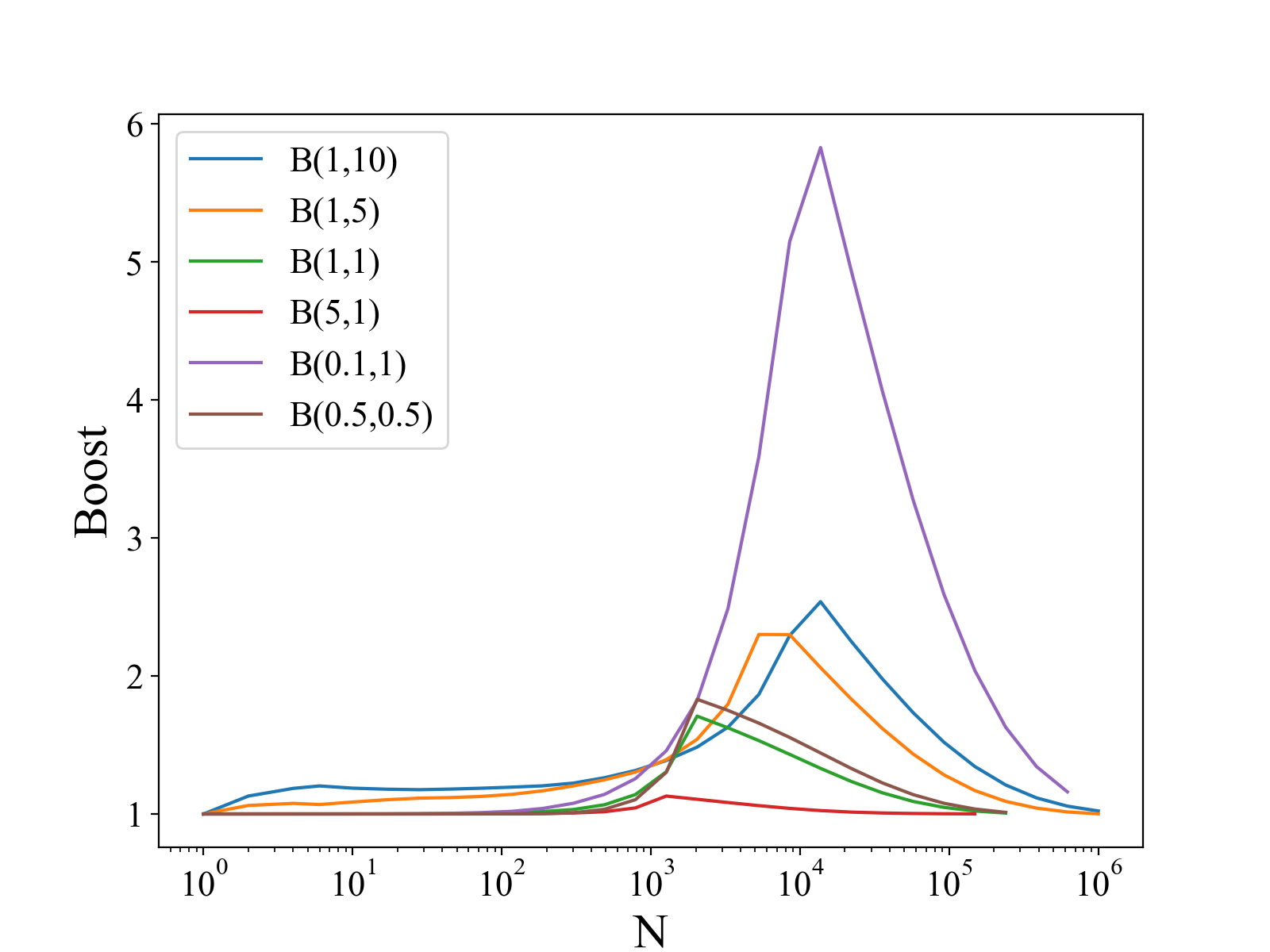}
         \caption{}
         \label{fig:Betaboost}
     \end{subfigure}
    \caption{Boost in performance as a function of resources $N$, where Boost is defined as the ratio of expected throughput provided the optimal allocation to the expected throughput provided the naive allocation. Results shown for synthetic distributions sampled from both the delta distributions (a) and the beta distributions (b).}
\end{figure}

We first recognize from the blue curve in Figure \ref{fig:Deltaboost} that the trivial probability distribution where all tasks are of equal probability $(p_i=p)$, corresponding to task probabilities sampled from a single delta distribution, obtains unit boost in performance compared to naive scheduling throughout the entire range of $N$.  This was expected given that when all probabilities are equal, the throughput is maximized for a uniform allocation of resources.
The natural extension to this trivial case of uniform task-probabilities is the case of binary probability values, where tasks are assigned one of two probabilities: $p_a$ and $p_b$ (where $p_a>p_b$). Such synthetic task probabilities are sampled from a composition of delta distributions, e.g., constructing a list of containing certain ($p_a=1$) and speculative ($p_b<1$) tasks. An example of a step distribution arising in practice might be an application which identifies a certain number of tasks as provably necessary (and hence for which $p_a=1$), and a certain number of speculative tasks, which are assigned a generic probability $p_b$. Synthetic task probability distributions were sampled for four different values of speculative task probability ($p_b = \{0.5, 0.1, 0.01, 10^{-10}\}$). It is seen in Figure \ref{fig:Deltaboost} that the boost obtained through the optimal allocation varies inversely with $p_b$ and saturates as $p_b$ approaches zero. 


Take for example the resource allocation of $N=10,000$ to a probability distribution characterized by $p_a=1$, $p_b=0.01$. The naive allocation distributes resources evenly across all possible tasks, producing an expected throughput of roughly 2.3. This is in stark contrast to the optimal allocation, which concentrates resources only to $p_a$ tasks. As a result, the optimal allocation executes fewer tasks, but does so yielding a higher expected throughput of roughly 16.8. The optimal allocation therefore results in a substantial boost in performance, producing nearly 7.3 times the expected throughput of the naive allocation. We note that the boost also affects only an intermediate range of values of $N$, as, in both the small and large $N$ limits, the optimal and naive allocation are identical.

The value of $p_b$ relates to the potential boost obtained as it affects the sampled task probability distribution in two key ways: $1)$ The smaller values of $p_b$ present steeper decays in the probability distributions as they cover a larger range in values. The naive allocation struggles to handle a large range in values as its allocation is uniform, meanwhile the optimal allocation is specifically tailored to the individual probabilities of each task. $2)$ Because the values ($p_a$, $p_b$) are sampled in a 1:9 ratio, a smaller value of $p_b$ implies that more tasks will be sampled before reaching the $\sum p_i=1000$, and thus the distribution of tasks will have a longer tail. This, again, is not handled well by the naive allocation as the high $p=p_a$ tasks will receive the constant $w_\mathrm{const}=1$ allocation unless $N$ is such that all tasks can be executed, at which point $w$ will start to increase uniformly. The longer the tail in the distribution, the more resources are needed before the naive allocation will increase the uniform chunk size, and hence the throughput. 

These results illustrate the intuitive idea that ignoring the task probabilities and invoking a uniform allocation often involves running lower probability tasks with resources that could be better spent increasing the allocation to higher probability tasks. These two key features (steep decay and long tail) are particularly detrimental to the performance of the naive allocation scheme. 

\begin{figure}
    \centering
    \includegraphics[width=.9\linewidth]{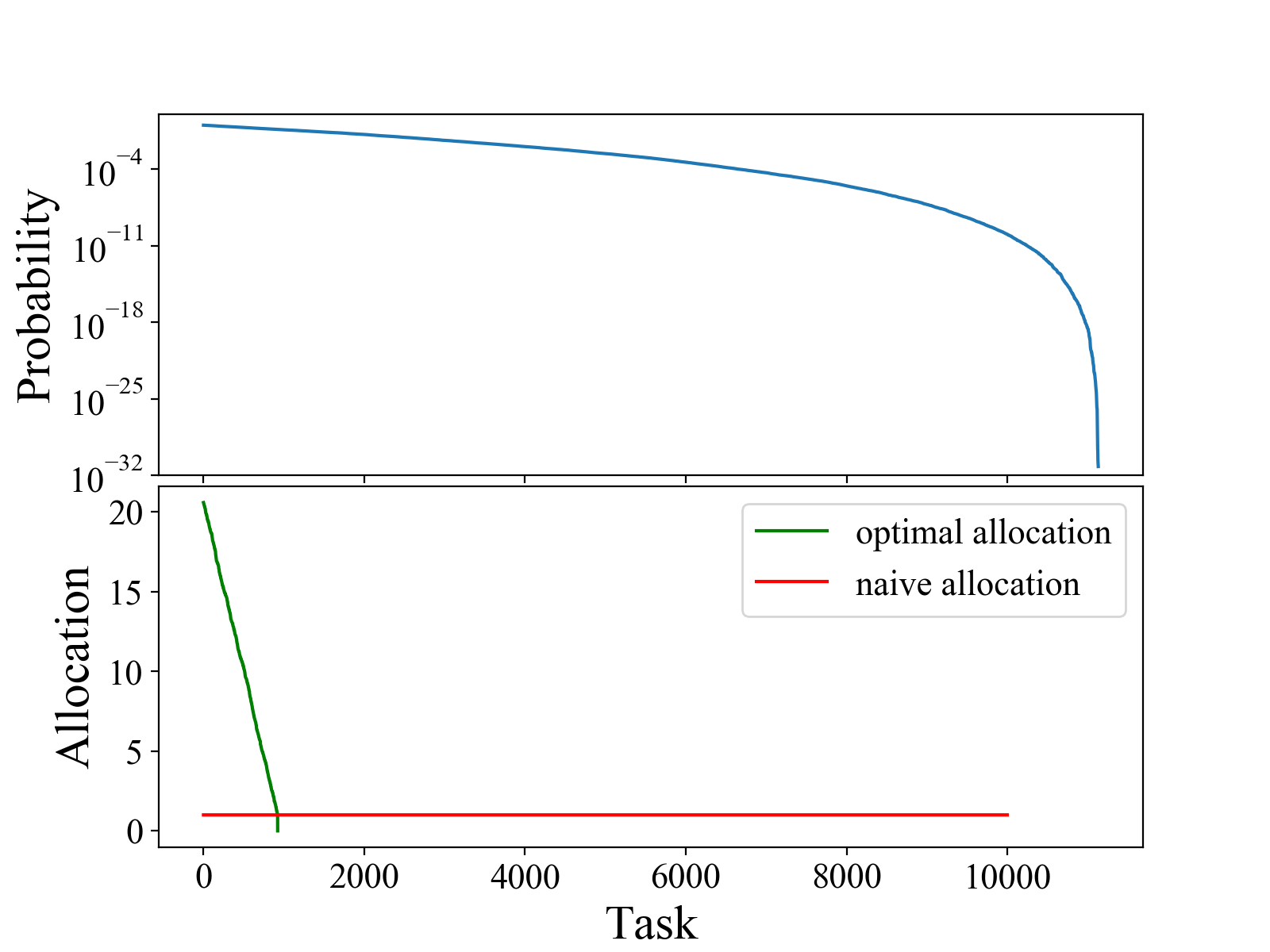}
    \caption{Top: Task probability distribution sampled from B$(0.1,1)$ distribution. Bottom: Allocation of $N=10,000$ resources among tasks.}
    \label{fig:10kBeta}
\end{figure}

Considering the task probability distributions sampled from the Beta probability density, one can see that this rule of thumb is upheld. Figure \ref{fig:10kBeta} illustrates allocation solutions for $N=10,000$ given a task probability distribution sampled from B$(0.1,1)$. The sampled probability distribution consisted of 11,132 tasks and spanned a range of probabilities from $p\sim1$ to $p\sim10^{-32}$. The naive allocation allocated resources uniformly, executing $10,000$ tasks with $w=1$. This resulted in an expected throughput of just over 2. The optimal allocation provided resources in greater chunks to fewer tasks. It only executed 923 tasks, but did so yielding an expected throughput of nearly 12, providing a boost in expected throughput of nearly 6 times the naive allocation.

We see here again how the long tail and steep decay of the task probability distribution meant that the naive solution would allocate resources to low probability tasks that contribute little to the expected throughput. It is instead optimal to allocate additional resources to those higher probability tasks, running fewer tasks but generating a higher expected throughput. 

The maximum attainable boost one could possibly obtain can be determined by the following analysis. Consider a task probability distribution consisting of $N_a$ tasks of probability $p_a=1$ and $N_b$ tasks of probability $p_b=\epsilon$. The optimal allocation would divide resources among those $N_a$ tasks, ignoring the $N_b$ tasks. This would yield an expected throughput of $N_a/T(N/N_a)$. The naive allocation would spread resources among all tasks, yielding an expected throughput of $N_a/T(1)+N_b \epsilon/T(1)$ if $N$ was sufficiently large such that $N=N_a+N_b > w_\mathrm{max} N_a$. In the case where $N_b \epsilon <<1$ this expression would simplify and a trivial relation for the boost in expected throughput would result as the ratio $T(1)/T(N/N_a)$. This expression is maximal when $N/N_a=w_\mathrm{max}$. Thus, the maximum attainable boost one could obtain is equal to $T(1)/T(w_\mathrm{max})$, which for our application was roughly 25. 


\subsection{Constant $w$}

\begin{figure}
    \centering
      \begin{subfigure}[b]{.49\textwidth}
         \centering
         \includegraphics[width=\linewidth]{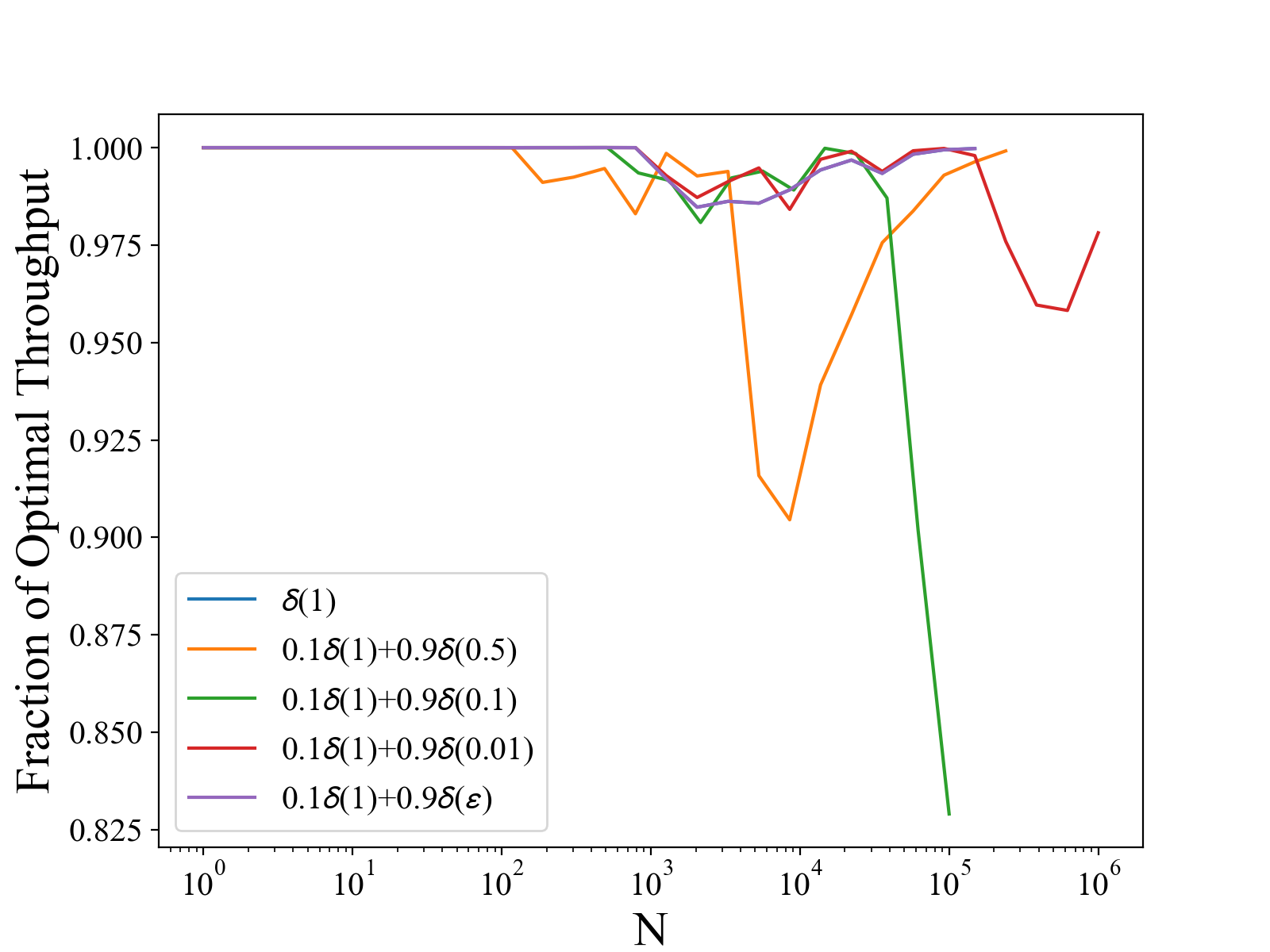}
         \caption{}
         \label{fig:FracoTPDelta}
     \end{subfigure}
     \hfill
   \begin{subfigure}[b]{.49\textwidth}
         \centering
         \includegraphics[width=\linewidth]{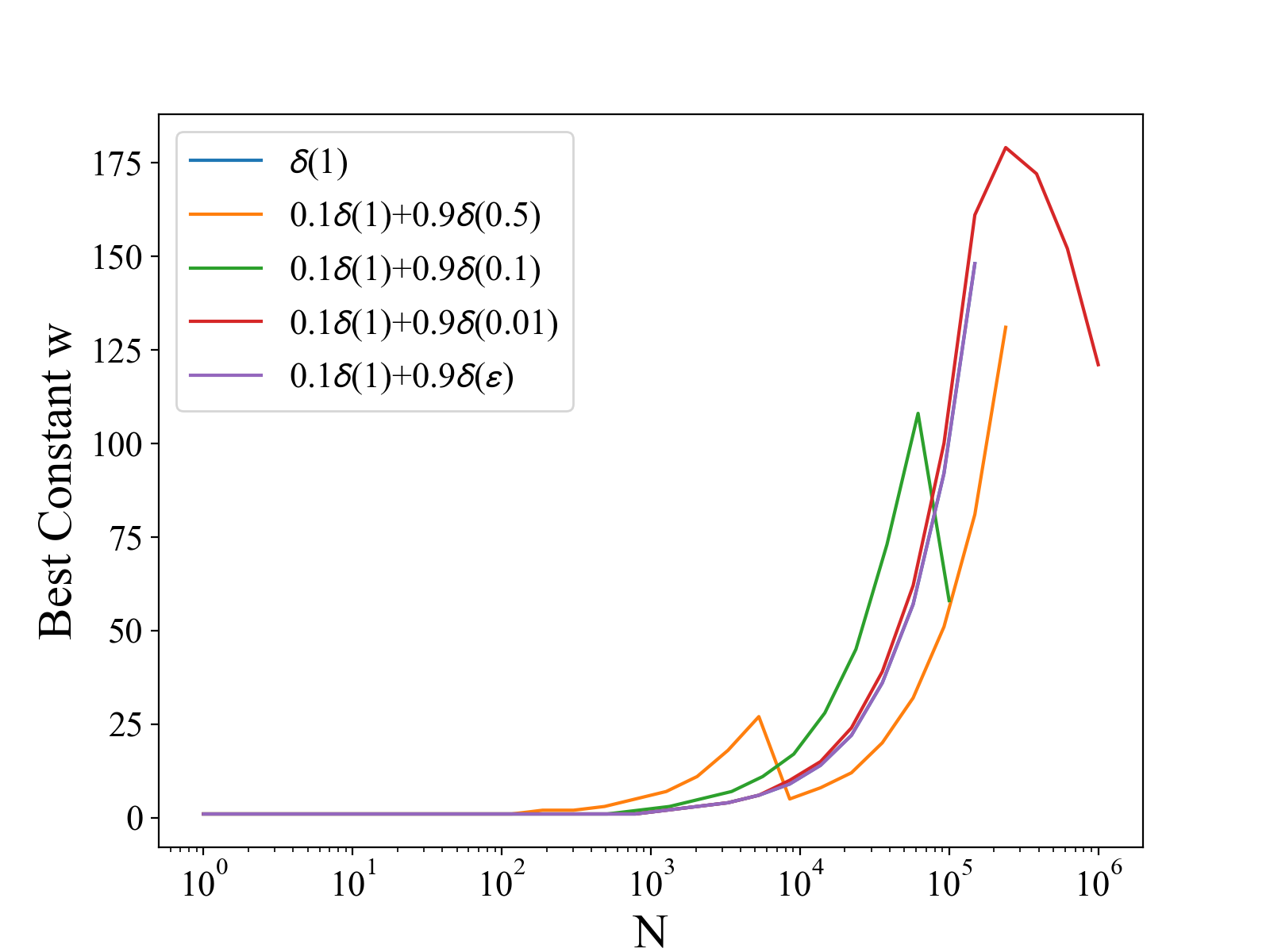}
         \caption{}
         \label{fig:BestConDelta}
     \end{subfigure}
     \begin{subfigure}[b]{.49\textwidth}
         \centering
         \includegraphics[width=\linewidth]{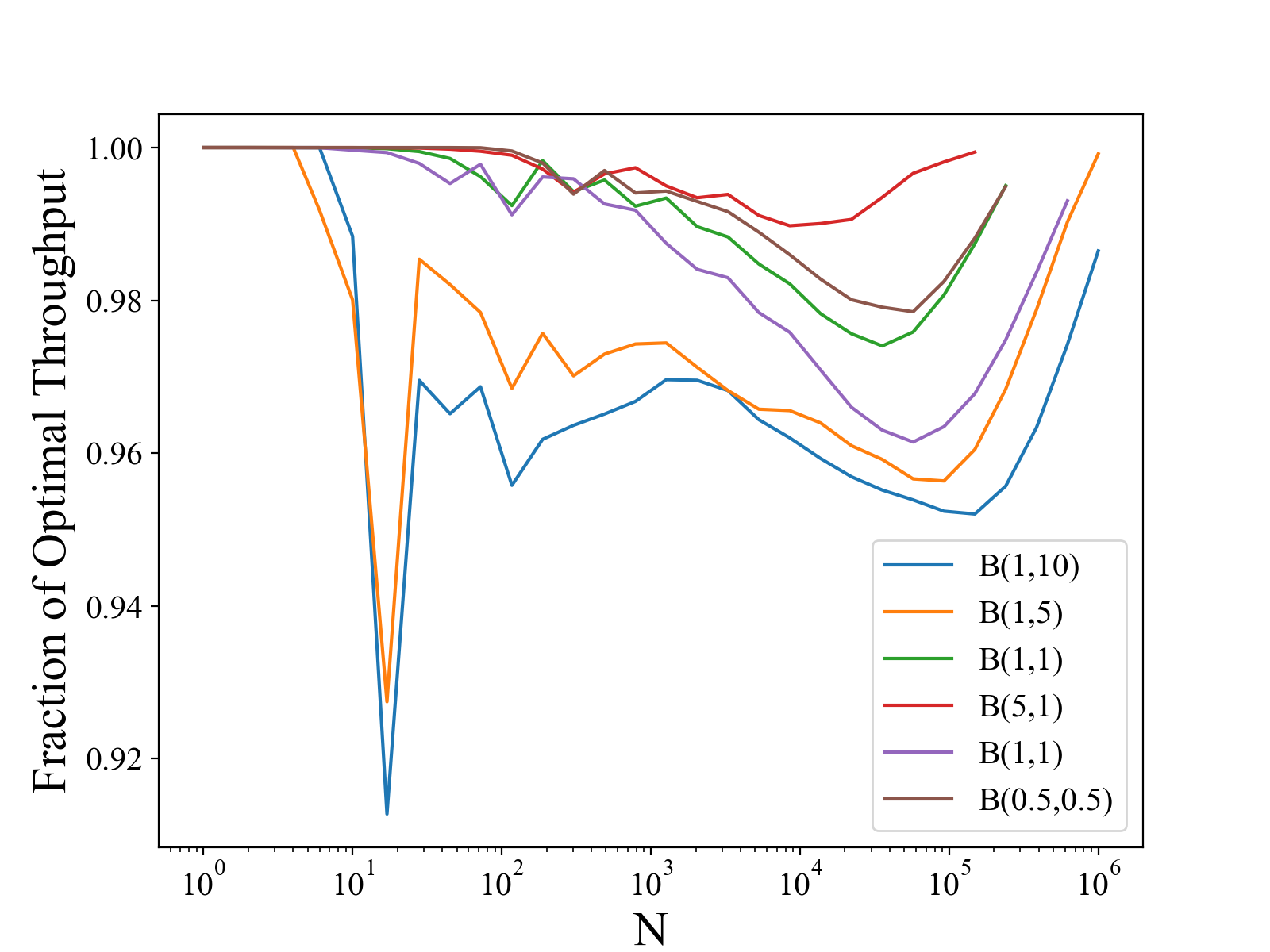}
         \caption{}
         \label{fig:FracoTPBeta}
     \end{subfigure}
     \hfill
     \begin{subfigure}[b]{.49\textwidth}
         \centering
         \includegraphics[width=\linewidth]{fig8c.png}
         \caption{}
         \label{fig:BestConBeta}
     \end{subfigure}
    \caption{On left, fraction of optimal throughput which can be achieved with a constant allocation for the Delta sampled distributions (\ref{fig:FracoTPDelta}) and Beta sampled distributions (\ref{fig:FracoTPBeta}). On right, the corresponding value of $w$ required to obtain this fraction of optimal throughput for the respective Delta sampled (\ref{fig:BestConDelta}) and Beta sampled (\ref{fig:BestConBeta}) distributions.}
    \label{fig:boost}
\end{figure}

The assortment of synthetic distributions surveyed above shows a diverse range of optimal task allocations and the corresponding boost compared to the naive scheduling strategy. These task allocations are guaranteed to provide the greatest expected throughput for a given distribution, at the cost of increased code complexity. One may instead consider a simpler approximate solution where each executed task is provided the same allocation, but this allocation is allowed to differ from the naive strategy. Certainly this simplified scheme would be suboptimal, but it is unclear by how much.  In the trivial case of constant probability, for example, the optimal allocation was a constant allocation. The same was true for the step distributions when resources are limited. In fact, it is often the case that a constant allocation can achieve a throughput close to that of the optimal allocation. Figures \ref{fig:FracoTPBeta} and \ref{fig:FracoTPDelta} show what fraction of the optimal expected throughput can be achieved when an optimal constant allocation is provided for each of the distributions surveyed above. For most values of $N$ there exists a constant value of $w$ which can provide upwards of 90\% the expected throughput that the optimal allocation would yield. This is, however, not always the case. When the task probability distribution possesses a major discontinuity (as seen in the step distributions), there exists a range of N values where even the best constant value is largely suboptimal. 

Furthermore, as seen in figures \ref{fig:BestConBeta} and \ref{fig:BestConDelta}, the value of the best constant $w$ can greatly vary depending on the resources available and on the specifics of the task probability distribution. Furthermore, in a setting where the task probability distribution is dynamic and/or context dependent, the precise value of the best constant allocation will change in time. This presents a major difficulty in assigning a single constant value of $w$ that will be allocated to each executed task; what may be a “good” value of $w$ at one point in time might be a poor value sometime later. To maintain an expected throughput that is near optimal for an evolving task probability distribution, one would have to consistently tune the constant value of $w$. Devoting this effort to maintain a near-optimal solution is uneconomical as one could ensure the truly optimal solution with similar efforts.

\subsection{ParSplice Simulator}

In order to assess the potential performance gains accessible with this new approach without the considerable time investment required to rewrite the ParSplice production code, we instead chose to make use of a simulator; a strategy which has proved beneficial \cite{spectad,tadsim} in developing Accelerated Molecular Dynamics \cite{perez2009accelerated} methods, to which ParSplice belongs. The ParSplice Simulator (ParSpliceSIM) was designed to directly mirror the logic of the actual ParSplice code with the exception that dynamics are generated from a user-specified Markov Chain that can be used to statistically sample segment endpoints, rather than using computationally expensive MD calculation \cite{garmon2020exploiting}. ParSpliceSIM is therefore computationally light, simple, and runs in serial. As the matter of interest is measuring the potential improvements in performance, rather than obtaining correct atomistic trajectories, the ParSpliceSIM provides an ideal framework for testing the proposed ParSplice variant described in this work.  

ParSpliceSIM was used to compare the performance of the existing VE method to the newly developed MaxP method, which was then further enhanced to allow the periodic pausing of segments and reallocation of resources following the optimization procedure described above. The effect of these successive developments are shown on a range of different model systems to illustrate the expected gains in performance. As to separate the effect of the current work from that of intrinsic model uncertainty, it was assumed that the MM created on-the-fly by ParSplice would provide a sufficiently accurate representation of the dynamics, and was instead replaced with the actual underlying Markov Chain within the ParSpliceSIM. The metric of performance in evaluating the different methods is the amount of pseudo-MD spliced for a given wall clock time (WCT), which is a direct measure of the scientific value of a simulation.

The following ParSpliceSIM results were generated using an assortment of Markov Chains with varying state connectivity. Each state within a particular Markov Chain was endowed with a $\rho_{ii}$ probability of not escaping the current state, and had an equally likely probability of transitioning to one of it’s $K$ neighbors $(\rho_{ik}=(1-\rho_{ii}) / K)$ with $K$ defined by connectivity. Periodic boundaries were implemented to ensure each state within a particular Markov Chain had the same state connectivity. To mimic the environment of a true atomistic simulation, the number of states in a particular Markov Chain was set to 8000 such that the vast majority of state-space remained un-visited by the trajectory throughout the duration of the simulation. 

In order to evaluate the true potential of the developed methods, each simulation was provided a resource allotment $N$ that was many times greater than the expected number of segments needed to escape ($\left\langle n_\mathrm{escape} \right\rangle$) from a state. This is the regime of greatest interest as it is where speculation significantly affects the efficiency of ParSplice. 
When the resources are not greater than $\left\langle n_\mathrm{escape} \right\rangle$ all resources can be allocated to the task(s) of building in the current state and will be amortized with high probability.
Prior to the current work, ParSplice attempted to best utilize those additional resources (those which were not likely to be needed in escaping the current state) to speculate on where the trajectory was likely to go next. The main purpose of this section is to assess whether further efficiency gains are possible by dynamically assigning resources based on expected utility.

In what follows, each simulation was carried out with assigned $\rho_{ii}=0.99$, corresponding to $\left\langle n_\mathrm{escape} \right\rangle=100$, and a resource allotment of $N=5000$. With an $N$ being $50$x greater than $\left\langle n_\mathrm{escape} \right\rangle$ ParSplice can trade-off resources in order to obtain an escape from the current state faster. This tradeoff would be beneficial in cases where speculation was futile, but could be poor in cases where accurate speculation were possible. To illustrate this effect, results are shown for three different Markov Chain toy models of increasing connectivity: 1D representing dynamics on a line, 3D representing dynamics on a cubic lattice, and fully-connected, where each state is connected to every other state. The greater the connectivity, the more difficult it is to speculate on the trajectory's future as the possible paths exhibit exponential branching. Conversely, when the connectivity is low speculation can be quite accurate. While these  models of state connectivity are much simpler than what would be observed in actual applications they do however provide relevant information and general guidelines on the potential performance of the method in different scenarios. 

These three toy models present a good assortment for testing the new methods. The 1D model provides rather-predictable dynamics for which speculation will be quite fruitful. The 3D model presents dynamics which are somewhat less predictable, and where effective forecasting will likely be limited to within a small neighborhood from the current state. Finally, speculation is futile for the fully-connected model where the number of branching paths is immense. 

\begin{figure}
    \centering
    \includegraphics[width=\linewidth]{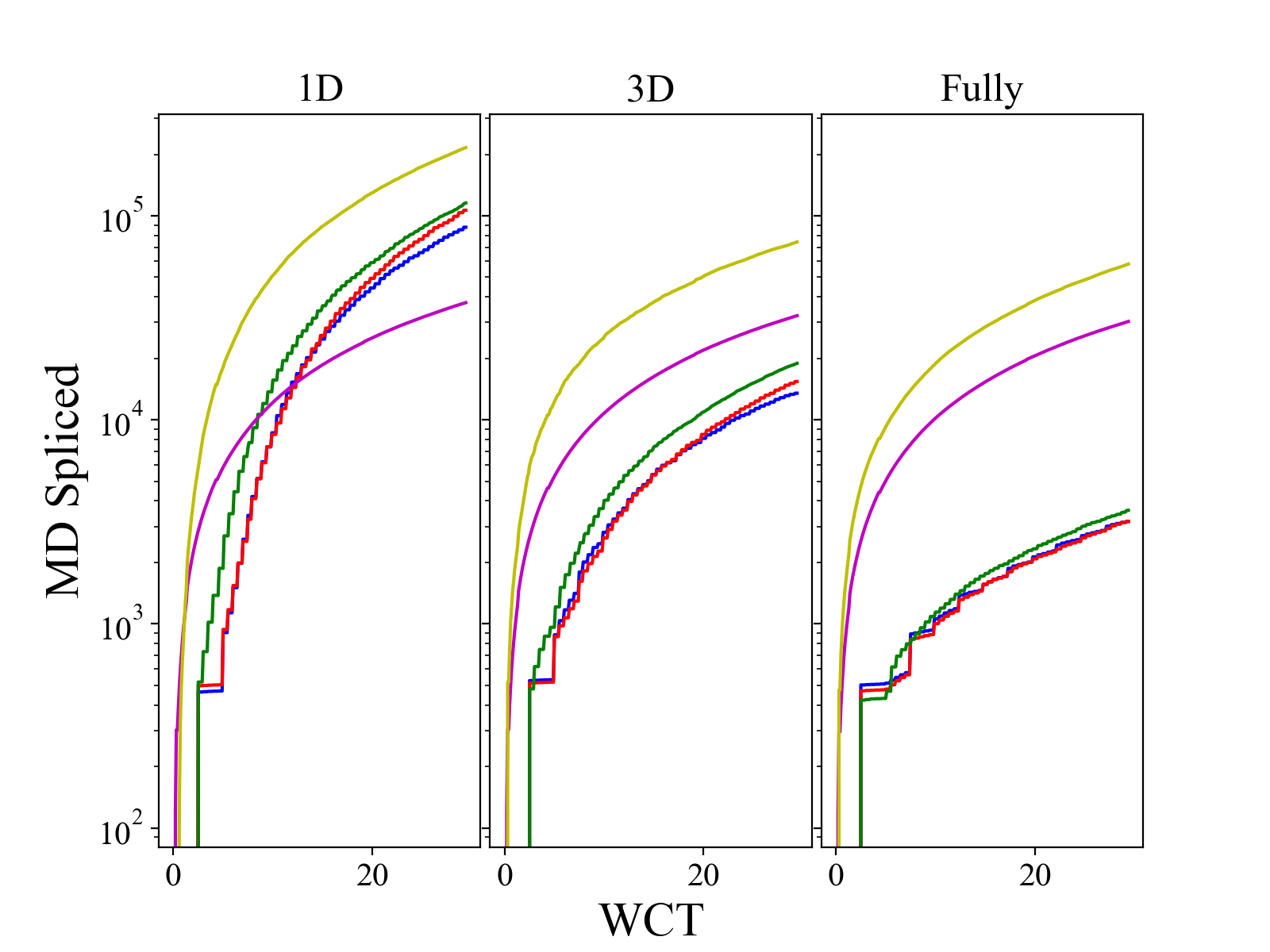}
    \caption{ParSpliceSIM results for the 1D, 3D, and fully-connected toy models showing pseudo-MD spliced as a function of WCT. Each panel displays performance of VE in blue, MaxP in red, MaxP($w_\mathrm{const}$) in green, MaxP($w_\mathrm{max}$) in maroon, and MaxP($w^*$) in yellow. The results shown represent an average of roughly 500 independent simulations conducted for each method on each model.}
    \label{fig:SIM}
\end{figure}

Performance results for each of the toy models are shown in Figure \ref{fig:SIM}, displaying the pseudo-MD spliced as a function of WCT. Each subfigure shows the results of five different methods for its particular model. The different methods consist of the existing VE formalism, the newly introduced MaxP formalism, and MaxP with preemption and restarts. The last method is shown implementing three distinct allocation policies: 
1) The (naive) maximum throughput allocation; distributing resources evenly $(w=w_\mathrm{const})$ to execute the most tasks at the highest throughput, thus producing the maximum number of segments for a given $N$. 
2) The minimum time allocation; distributing resources evenly to execute tasks with the maximum allocation (as defined previously, $w=w_\mathrm{max}$) thus producing $N/w_\mathrm{max}$ segments as quickly as possible. 
3) The (optimal) maximum expected throughput allocation; distributing resources according to how likely tasks are to be spliced onto the state-to-state trajectory, thereby balancing the tradeoff between time and throughput to produce the most spliced segments as quickly as possible. 

The first thing to note in Figure \ref{fig:SIM} is the small but appreciable increase in performance that results from transitioning from the VE to the MaxP formalism. While MaxP was introduced to allow for the implementation of our derived methods, it is worth noting that the transition does not come at the cost of performance, to the contrary.

 Further improvement resulting from the ability to pause and reschedule segments can be substantial depending on the topology of the state space. As was discussed previously, the 1D toy model presents very limited connectivity, therefore corresponding to a system which is highly susceptible to speculation. As a result, the distribution of task probabilities will decay slowly, and the balance between running a few tasks very quickly and maximizing the overall task-completion throughput will be more heavily skewed toward throughput. This is seen from the 1D results in Figure \ref{fig:SIM} as the maximum throughput allocation outperforms the minimum time allocation (which even under-performs the standard VE and MaxP) by a factor of three. However, even in a highly predictable system like the 1D toy model, the balance between time and throughput is not completely one-sided. This is seen as the optimal allocation (which aims to maximize the \textit{expected} throughput, or segments spliced) is able to further improve performance by a factor of two as it strikes the optimal balance. Overall, the implementation of our derived methods applied to the 1D model are able to more than double the pseudo-MD spliced over the same WCT as compared to the standard VE method.

The 3D model presents a slightly different picture as long-time speculation is somewhat difficult due to the increased connectivity, yet short-time speculation can still be profitable, thus this model requires a more delicate balance between throughput and execution time. This is seen from the 3D results in Figure \ref{fig:SIM} as now the minimum time allocation outperforms the maximum throughput allocation by over 50\%. The optimal allocation adapts to the new model and achieves the best performance, more than doubling the efficiency of the the minimum time allocation strategy and providing nearly a six-fold improvement as compared to the standard VE method. 

Lastly, the fully-connected model is considered, for which speculation is futile and escaping from the current state as quickly as possible is the only sound strategy. As expected, the fully-connected results in Figure \ref{fig:SIM} show how the minimum time allocation greatly outperforms the maximum throughput allocation by nearly an order of magnitude. However, the optimal allocation is able to further improve performance by nearly doubling the throughput achieved over the simulated times shown here. Although the minimum time allocation utilizes resources to generate segments as quickly as possible, it does not achieve the desired result of escaping from the current state as quickly as possible. This is because the number of segments it produces is likely insufficient to escape the state, i.e less than $\left\langle n_\mathrm{escape} \right\rangle$. It is actually better to generate more segments (greater throughput) at a slightly slower rate (but higher efficiency) such that a sufficient number of segments to escape from the current state are generated. Overall, our derived method applied to this toy model of greater connectivity enabled nearly twenty times the pseudo-MD to be spliced over the same WCT as compared to the standard VE method.

\begin{figure}
    \centering
    \includegraphics[width=.8\linewidth]{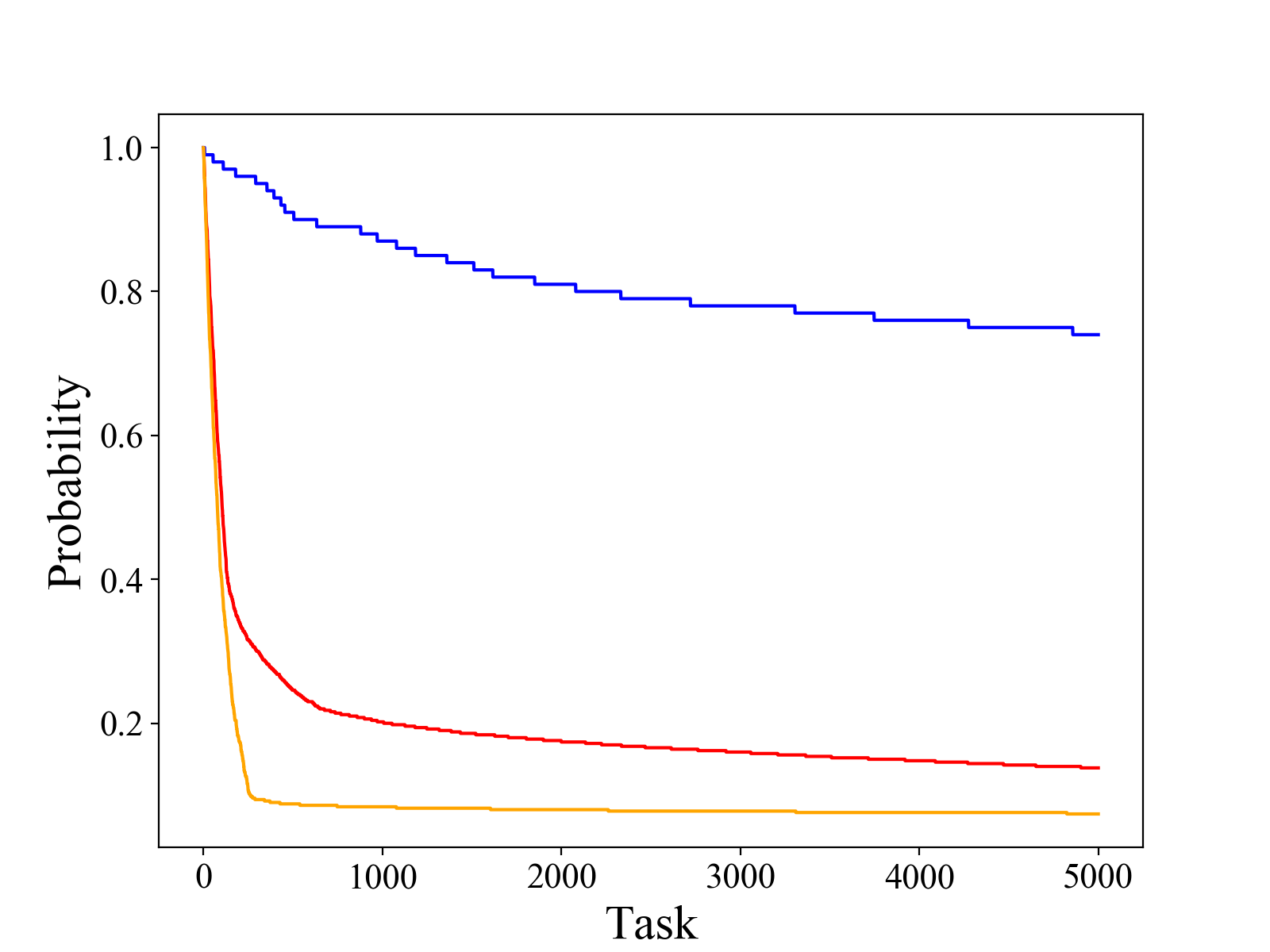}
    \caption{Initial task probability distributions taken from simulations on the 1D (blue), 3D (red), and fully-connected (orange) toy models. These task probability distributions were constructed at the start of the simulation; having no contribution from the then-empty database of stored segments, and are therefore a reflection of the state connectivity.}
    \label{fig:single}
\end{figure}

The resulting improvement of our derived methods, as compared to the existing VE scheduling method, showed a nearly 2.5x, 6x, and 20x boost in performance for  1D, 3D, and fully connected toy models, respectively. These ParSpliceSIM results can be better understood by analyzing the task probability distributions that are characteristic of each toy model. Figure \ref{fig:single} shows an example initial probability distribution for each of the toy model systems as was constructed by the MCMaxP procedure described in appendix A. One can see how the task probabilities for the 1D model exhibit a very gradual decay over the first 5,000 tasks down to $p\sim0.8$. This reflects  the limited state connectivity which makes speculation fruitful. The 3D task probabilities exhibit a very steep decay over the first $\sim$100 tasks (corresponding to an escape from the current state), followed by a more gradual decay over the following 600 tasks (corresponding to an escape from the 6 neighbors of the current state), followed by a more gradual decay out to 5,000 tasks. 
Overall, task probabilities remain non-negligible out to 5,000 tasks with long tail around $p\sim0.2$. Considering the fully connected model, one can see a sharp decay in probability corresponding to the first escape from the current state, after which the probability of tasks drops down to $p<0.1$. One can see how these results in performance generally adhere to the inference made while studying the synthetic distributions, i.e. performance gains are largest when the probability distribution exhibits steep decays and long tails.

\begin{figure}
    \centering
    \includegraphics[width=.8\linewidth]{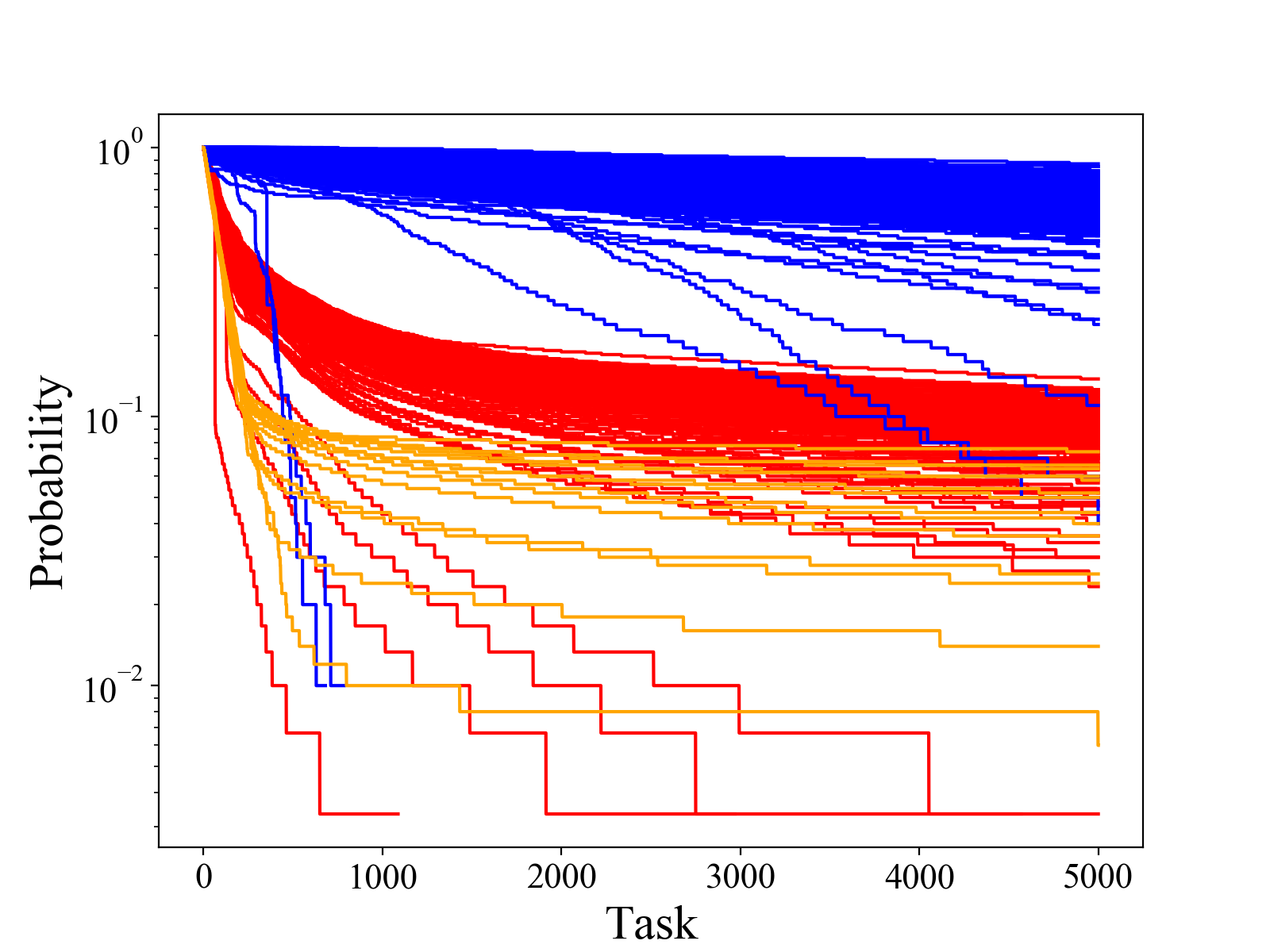}
    \caption{All task probability distributions generated during a single simulation on the 1D (blue), 3D (red), and fully-connected (orange) toy models.}
    \label{fig:all}
\end{figure}

Figure \ref{fig:all} shows the evolution of task probability distributions which were sampled throughout a simulation for each of the toy model systems. This occurs for two reasons: 1) The effect of the database; the stored segments which have been generated but not yet consumed by the trajectory play a role in the MCMaxP sampled trajectory and therefore affect which segments are expected to be ``needed” by a future trajectory. And 2) the time horizon over which the MCMaxP trajectory is sampled; a greater time horizon corresponds to a higher likelihood that a particular segment will eventually be spliced into the trajectory, thus resulting in a shallow decay of the probability distribution.
The dynamic nature of the probability distributions is quite relevant as it pertains to a changing allocation which must be maintained in order to achieve optimal performance. Recognizing this fact, one may consider the difficulties of maintaining the optimal allocation and note how the frequency at which segments are paused and resources are reallocated can have a large impact on performance. Constantly maintaining the optimal allocation involves pausing and reallocating resources whenever any new information is available, which is not always feasible. In practice, the user may choose to pause and update the allocation at some fixed interval, which can be tuned for the user’s particular system. In the present study, the allocation was updated whenever a segment was completed. This can be thought of as the most aggressive scenario that will produce the upper bound in performance. In the ParSpliceSIM results above this condition could be relaxed to update whenever a segment contained a transition as the true underlying Markov Chain was being used rather than developing a model from segment data, and therefore segments which did not yield escapes would have no effect on the current MCMaxP-constructed task probability distribution. In a true ParSplice simulation however, each segment would contribute to the development of the statistical Markov Model being produced on-the-fly and thereby (possibly) change the statistics of the MCMaxP sampled trajectories, hence affecting the task probability distribution. 

The question of when it is appropriate and necessary to pause and reallocate resources is not easy to quantify. In general, the user would want to do so whenever the task probability distribution substantially changes. In the example of ParSplice, this would certainly occur whenever the current state of trajectory changes as all tasks and probabilities are generated from the MCMaxP procedure and are thus conditional on starting in a particular state. The distribution could also change without the trajectory changing state, however, when unused segments which contain transitions are stored in the database and can contribute to the MCMaxP sampled trajectories. For this reason the ParSpliceSIM results were aggressively updated whenever a segment yielding a transition was returned to the database. The question becomes even more fuzzy for a real ParSplice simulation which develops it’s statistical model on the fly. Any segment which changes the model in a substantial way will likely change the MCMaxP sampled trajectories and thus affect the sampled probability distribution. It is not easy to systematically catch changes of this type. A single segment will likely have a negligible effect on the model, but cannot be ignored outright as enough of these negligible changes can account for a significant effect. In addition, a segment which drastically changes the model but does so far from the current state of the trajectory will have little effect on the sampled task probability distribution and therefore does not warrant pausing and reallocating resources.  In practice, it is pragmatic to periodically pause and update the allocation at some fixed interval, which is chosen so as to limit scheduling and pausing/restarting overhead.

\section{Conclusion}
The advent of exascale computing platforms will be accompanied by a need for specially designed software and algorithms that are capable of utilizing the large availability of resources simultaneously. As maintaining strong-scalability on such platforms will be quite difficult, the use of speculative task-based paradigms are promising; enabling higher concurrency and improved scaling. In this work, we derived the optimal allocation of resources for task execution in this speculative setting. The utility of this approach was then assessed on assortment of synthetic task probability distributions, comparing the expected throughput of our derived optimal allocation of resources to more naive allocation policies. While a uniform allocation of resources can often be found to produce a nearly optimal expected throughput, it was shown that determining the particular value for the constant allocation size is in practice just a difficult as computing and employing the optimal allocation. 

A dynamic setting was then considered where task probabilities were influenced by some underlying variable (state, context, time, etc.) and were therefore changing throughout the run-time of the application. 
This setting was explored by examining the effect of our derived methods applied to a specific scientific application, ParSplice, which operates in this domain. In order to implement our methods, we first had to design a new application-specific technique for accessing the speculative probability that potential tasks would be useful. This technique not only allowed for our derived methods to be implemented, but was also shown increase the performance of the scientific application. The potential gains in performance resulting from our derived methods were assessed through the use of a simulator. While the boost achieved varied with physical system  (ranging from 2.5x to 20x), it was found to be greatest when the system of study was most complex; resulting in lower speculative task probabilities and a greater ability to leverage the trade-off between throughput and time. By considering the speculative task probabilities, the optimal balance could be struck to produce the maximum rate of expected throughput. This novel optimization scheme stands to improve performance of speculative task-based applications, particularly when run at large computational scales.

\section*{Acknowledgments}
AG was supported by the US Department of Energy Office of Science Graduate Student Research (SCGSR) program. The SCGSR program is administered by the Oak Ridge Institute for Science and Education (ORISE) for the DOE. ORISE is managed by ORAU under contract number DE-SC0014664. All opinions expressed in this paper are the author’s and do not necessarily reflect the policies and views of DOE, ORAU, or ORISE. VR was supported by the Advanced Simulation and Computing Program (ASC) and DP was supported by the Exascale Computing Project (17-SC-20-SC), a collaborative effort of the US Department of Energy Office of Science and the National Nuclear Security Administration. Los Alamos National Laboratory is operated by Triad National Security LLC, for the National Nuclear Security administration of the U.S. DOE under Contract No. 89233218CNA0000001. We graciously acknowledge computing resources from the Los Alamos National Laboratory Institutional Computing (IC) program, and insightful discussions with David Aristoff and Mouad Ramil.

\begin{appendices}

\section{Maximum-Probability (MaxP)}

In the following, the probability that a candidate task (e.g, the generation of a segment starting in a given state) will be consumed as part of the calculation is derived in the context of discrete time Markov Chains, which is the natural setting for a trajectory composed of segments generated following the ParSplice prescription. In other words, the problem at hand is to compute the probability that a trajectory of a given length, sampled from this Markov Chain, would contain the generated segment. As will be shown, these probabilities can be evaluated analytically from the Markov jump process or approximated through a Monte Carlo sampling procedure. To clearly distinguish from any variables defined in the main text, we have chosen to express this derivation using a double-struck font. 

This derivation utilizes a Markov model $\mathbbm{M}$; a stochastic matrix whose elements $\mathbbm{p}_{ij}$ represent the probability of moving from state $i$ to state $j$, thereby governing the discrete Markov process. In the example of our scientific application (ParSplice) these are the probabilities that a segment starting in state $i$ will end in state $j$. Note that these probabilities encode the potential outcome of the task, not the potential task's usefulness, and are therefore distinct from task probabilities $p_i$ defined in the main text. The work detailed herein describes our method for extracting these latter probabilities $p_i$ from the former probabilities $\mathbbm{p}_{ij}$. 

We define $\mathbbm{f}_{ij}^{(n)}$ as the probability that the first passage from state $i$ to state $j$ takes exactly $n$ steps. This can be written recursively as

\begin{align*}
    \mathbbm{f}_{ij}^{(1)} & = \mathbbm{p}_{ij}^{(1)}=\mathbbm{p}_{ij} \\
    \mathbbm{f}_{ij}^{(2)} & = \sum_{k!=j} \mathbbm{p}_{ik} \mathbbm{f}_{kj}^{(1)} \\
    \mathbbm{f}_{ij}^{(n)} & = \sum_{k!=j} \mathbbm{p}_{ik} \mathbbm{f}_{kj}^{(n-1)}
\end{align*} 

We then also define $\mathbbm{f}_{ii}^{(n)}$ to be the probability that the first return to state $i$ upon leaving state $i$ takes exactly $n$ steps, which can similarly be written recursively as

$$\mathbbm{f}_{ii}^{(n)}=[\mathbbm{M}^n]_{ii}-\sum_{k=1}^{n-1} \mathbbm{f}_{ii}^{(k)}[\mathbbm{M}^{n-k}]_{ii}$$
Where $[\mathbbm{M}^n]_{ii}$ represents the $i,i$ element of the Markov Model $\mathbbm{M}$ raised to the power $n$. Then, summing over the index $n$ allows one to write the probability of a return to state $i$ sometime in the next $\mathbbm{N}$ steps:
$$\mathbbm{F}_{ii}^{(n)}=\sum_n^\mathbbm{N} \mathbbm{f}_{ii}^{(n)}$$

Lastly, let $\mathbbm{v}_{j}$ be the number of visits to state $j$, and $\mathbbm{P}_{i}(\mathbbm{X})$ denote the probability of $\mathbbm{X}$ conditional on starting in state $i$. The probability of making exactly $m$ visits to state $j$ over the next $\mathbbm{N}$ steps from the current state $i$ can then be expressed recursively as
\begin{align*}
    \mathbbm{P}_{i}(\mathbbm{v}_{j}=1|\mathbbm{N}) & = \sum_{k=1}^\mathbbm{N} \mathbbm{f}_{ij}^{(k)}[1-\mathbbm{F}_{jj}^{\mathbbm{N}-k}]\\
    \mathbbm{P}_{i}(\mathbbm{v}_{j}=2|\mathbbm{N}) & = \sum_{k=1}^\mathbbm{N} \mathbbm{f}_{ij}^{(k)} \mathbbm{P}_{j}(\mathbbm{v}_{j}=1|\mathbbm{N}-1) \\
    \mathbbm{P}_{i}(\mathbbm{v}_{j}=m|\mathbbm{N}) & = \sum_{k=1}^\mathbbm{N} \mathbbm{f}_{ij}^{(k)} \mathbbm{P}_{j}(\mathbbm{v}_{j}=m-1|\mathbbm{N}-m)
\end{align*} 

Summing over the index $m$ and subtracting from $1$ yields the probability of having more than $S$ visits to a state over a horizon of $\mathbbm{N}$ steps:
$$\mathbbm{P}_{i}(\mathbbm{v}_{j}>S|\mathbbm{N})=1-\sum_{m=1}^S \mathbbm{P}_{i}(\mathbbm{v}_{j}=m|\mathbbm{N})$$

Therefore, given a current state of the trajectory $i$ and the number of pending/unconsumed segments in state $j$, this prescription provides a means of extracting the probability that the next segment generated in state $j$ will be consumed into the trajectory over the finite time horizon $\mathbbm{N}$. Denoting the number of pending/unconsumed segments in state $k$ as $S_k$ allows the the probability of each potential task to be written as 
$$p_k=\mathbbm{P}_{i}(\mathbbm{v}_{k}>S_{k}|\mathbbm{N}),\forall k$$

Having this derivation in mind, we propose a new ParSplice scheduling scheme referred to as MaxP (maximum probability). In MaxP, the probability that each task will be spliced into the trajectory over a given finite time horizon is calculated. While computing these probabilities can be done analytically, it is often far more practical and efficient to do so via a Monte Carlo procedure, especially when the number of states is large.

This can be done in a similar fashion to VE where an ensemble of future state-to-state trajectories are sampled, accounting for the pending/unconsumed segments in the same way, but, instead of stopping when running out of segments, each trajectory continues until the preset time horizon, keeping track of how many segments would have to be generated in each state to reach said horizon. This ensemble is then used to calculate the probability that particular segments built in particular states are to be used by the state-to-state trajectory over the time horizon. ParSplice can then assign segments to be generated in decreasing order of probability, thus generating the segments which have the ``maximum probability" of being spliced. It can actually be shown\cite{MouadRamil} that the MaxP allocation scheme formally minimizes the expected number of database ``misses", i.e., the number of times splicing has to be interrupted because a segment that is required to move forward is not found in the database.

One may note that MaxP is substantially more expensive than VE for assigning an initial state to a single segment. While this is true, the ensemble average required by MaxP can be used to make state-assignments for a large number of segments all at once. This is compared to VE which can make one state-assignment for each virtual-trajectory. Furthermore, the VE process for assigning states to several segments must each be done in serial, meanwhile the ensemble trajectories used by MaxP can be generated in parallel. These differences can become quite significant as simulations scale to larger and larger machines, employing a greater number of MD instances. 
\end{appendices}

\bibliographystyle{unsrt}
\bibliography{references}
\end{document}